%% file: main.tex
\documentclass[11pt,5p,twocolumn]{article} 
\usepackage{authblk}
\usepackage{colortbl}
\usepackage{xcolor}

\usepackage{algorithm}
\usepackage{algpseudocode}

\usepackage{amsmath,amssymb,amsfonts}
\usepackage{graphicx}
\usepackage{textcomp}
\usepackage{mathtools}
\usepackage{microtype} 
\usepackage[inline]{enumitem}
\usepackage{multirow}
\usepackage{booktabs}
\usepackage{makecell}
\usepackage{url}
\usepackage{hyperref}
\usepackage{tikz}

\newcommand{\filledcircle}[1]{%
	\tikz[baseline=(char.base)]{
		\node[shape=circle, fill=black, text=white, inner sep=1pt] (char) {\textbf{#1}};
	}%
}

\usepackage[letterpaper, margin=0.7in, top=0.7in, bottom=1in]{geometry}

\begin{document}
\title{A Kubernetes-Native Request Router for Quality-Aware Inference Serving in the Computing Continuum}

\author[1]{Ignjat Karanovic}
\author[1]{Pantelis A. Frangoudis}
\author[2]{Ivan \v{C}ili\'{c}}
\author[2]{Ivana Podnar \v{Z}arko}
\author[1]{Schahram Dustdar}

\affil[1]{Distributed Systems Group, TU Wien, Austria}

\affil[2]{Faculty of Electrical Engineering and Computing, University of Zagreb, Croatia}

\date{}

\maketitle
\abstract{
	\input{./sources/abstract}
}

\input{./sources/intro}
\input{./sources/rwork}
\input{./sources/system}
\input{./sources/evaluation}
\input{./sources/conclusion}

\section*{Acknowledgment}
This work has been supported in part by the European Union's Horizon Europe research and innovation programme under grant agreement No. 101079214 (AIoTwin) and by the European Regional Development Fund (grant No. KK.01.1.1.04.0108, project IoT-Field).

\bibliographystyle{unsrturl}
\bibliography{./sources/bib}

\end{document}

%% file: sources/abstract.tex
We introduce \emph{Adaptive Score-based Routing Balancer (ASRB)}, a dynamic, score-based request routing mechanism for Kubernetes-based service deployments over the computing continuum. ASRB jointly considers infrastructure-level information, response time measurements, and application-level quality indicators, with a particular focus on serving Machine Learning (ML) workloads. For these workloads, ASRB balances requests over service instances deployed in the continuum, following service provider-defined policies encoded as weighted combinations of QoS criteria to flexibly address latency-accuracy trade-offs. To drive routing decisions and swiftly adapt to changes in the operating environment, ASRB monitors a range of runtime metrics across multiple system layers. To deal with the associated monitoring overhead, particularly important for large-scale deployments, it selectively and adaptively controls monitoring intensity without sacrificing on routing quality. ASRB is implemented without requiring any modifications to Kubernetes, making it straightforward to deploy and operate in existing cluster environments. Our testbed experiments demonstrate the versatility of ASRB: When tuned for latency reduction, it achieves at least 10\,ms lower mean response time compared with latency-oriented state-of-the-art routing mechanisms, while it achieves higher accuracy when this is prioritized through specific configurations, thus enabling flexible and operator-controllable trade-offs. At the same time, it attains reduced failure rates, higher responsiveness to changes in the operating environment, and up to $\sim$70\% less monitoring cost than relevant state-of-the-art solutions, at the potential expense of only a modest latency penalty in some configurations.

%% file: sources/intro.tex
\section{Introduction}
Modern IoT services are becoming increasingly data-driven, integrating Artificial Intelligence/Machine Learning (AI/ML) workflows as core components, and are characterized by multi-dimensional and often conflicting performance requirements. On the one hand, domains such as augmented reality~\cite{Younis20ar}, autonomous mobility~\cite{Turay22driving}, and real-time analytics~\cite{Chen23analytics} require consistently low response times to remain usable. The strive for bounded latency drives application deployment towards the edge of the computing continuum~\cite{Dustdar23dccs}, as cloud deployment can be prohibitive from a response-time perspective.
On the other hand, the prediction quality of the deployed ML models, typically expressed in terms of accuracy metrics, is crucial in domains like medical diagnostics~\cite{hicks2022evaluation} and safety-critical automation~\cite{PerezCerrolaza24safety}.

Achieving higher accuracy often incurs additional computational cost, which in turn translates to increased compute-induced latency and energy consumption. For example, deep neural networks for vision tasks, such as high-depth ResNet~\cite{he2016resnet} variants and Vision Transformers~\cite{Dosovitskiy21vit} deliver stronger prediction performance but require more processing time compared to lightweight architectures like MobileNet~\cite{Sandler18mobilenet} or to pruned and quantized model variants that reduce computational demand at the cost of accuracy~\cite{han2016deepcompression}. Cloud nodes may run such heavy models faster, but the additional network delay can offset these benefits. Conversely, edge nodes typically offer lower \emph{network} latency but may either host lightweight models with reduced accuracy or, if capacity allows, run more complex models with increased \emph{computation} latency due to resource limitations, thereby offsetting proximity benefits.

This reveals a trade-off between latency and prediction quality, which is important for service providers. Distributed service deployment over the computing continuum makes addressing this trade-off more challenging for the following reasons: (i) Compute nodes with diverse capabilities (e.g., resource-constrained edge devices vs. cloud servers) may be hosting replicated service instances; this implies inconsistent response times across these instances, while resource limitations can translate to service placement constraints. (ii) Similarly, different service instances come with end-to-end network paths of varying distance and thus latency. (iii) The volatile operating environment of the continuum requires intensive monitoring to keep a precise view of service and infrastructure state, which is critical input for orchestration decisions.

We address this particular trade-off in the context of inference serving over the continuum, focusing on the following orchestration problem: \emph{Given a number of service instances deployed over the continuum, each serving a given task (e.g., image classification) via an appropriate (but potentially different) ML model, what is the optimal service instance to handle a service invocation, considering response time and prediction quality objectives?}

We make the following contributions: \filledcircle{1} We present Adaptive Score-based Routing Balancer (ASRB), a request routing mechanism for ML inference serving, operating within a decentralized proxy architecture~(\S\,\ref{sec:system}) and allowing service providers to flexibly specify quality objectives within the latency-accuracy trade-off space~(\S\,\ref{sec:score-based-routing}). ASRB natively integrates with Kubernetes (K8s) and is open-source.\footnote{\url{https://github.com/Ignjat96/asrb-proxy}}
\filledcircle{2} We devise an adaptive monitoring scheme~(\S\,\ref{sec:monitoring}) that dynamically tunes monitoring intensity, considering the current fit of different nodes and their hosted service instances to attain given QoS goals, thus drastically reducing monitoring cost without significantly impacting ASRB's routing performance. 
\filledcircle{3} We demonstrate the flexibility of ASRB in addressing the latency-accuracy trade-off at reduced monitoring cost by evaluating it over a K8s-based testbed in a computer vision task~(\S\,\ref{sec:eval}).

%% file: sources/rwork.tex
\section{Related work}
\label{sec:rwork}
In Kubernetes-based environments, load balancing is typically handled by the built-in component kube-proxy, which distributes requests using basic round-robin or random strategies~\cite{kubernetes2023}. Solutions like service meshes (e.g., Istio\footnote{\url{https://istio.io/latest/docs/}}) offer customizable routing rules and observability, but often introduce operational overhead, making them less practical for lightweight or resource-constrained distributed edge deployments~\cite{Zhu23servicemesh}. QEdgeProxy~\cite{vcilic2024qedgeproxy}, upon which this work builds, maintains per-service dynamic pools of QoS-satisfying service instances which are candidates for routing, while proxy-mity~\cite{fahs2019proxy} routes based on a weighted combination of proximity (latency) and load balancing criteria. ProxyDWRR~\cite{wang2022proxydwrr}, on the other hand, treats CPU load as the sole decision factor. These works integrate well with K8s, but do not account for ML model-specific QoS attributes in routing decisions.

In contrast, ASRB unifies latency, model accuracy, and resource metrics in its routing strategy, enabling richer QoS trade-offs. Unlike centralized ProxyDWRR scheduling, ASRB uses decentralized proxies, whose routing decisions introduce negligible overhead on the ``hot path'' (i.e., per request). This contrasts approaches based on (Deep) Reinforcement Learning~\cite{santos2024towards, Karagiannis23, tang2020deep}, which, state management and training overheads aside, may require neural network execution per request. Nevertheless, ideas from RL, such as adapting parameters based on long-term QoS feedback, could complement ASRB. We have applied such concepts in another line of work~\cite{Cilic25mab}, albeit with pure latency orientation.

Notably, latency and accuracy trade-offs are addressed in model selection systems. MDInference~\cite{Ogden20mdinference} hosts a fast low-accuracy model on-device and a set of high-accuracy models in the cloud, and duplicates requests to ensure bounded response time, at the expense of energy due to duplication. RAMSIS~\cite{Mendoza24ramsis} exploits periods of low load to direct requests to higher accuracy models. However, it requires a central controller over which requests are routed. Jellyfish~\cite{Nigade22jellyfish} jointly instruments clients to adapt the size of their input data and adaptively maps clients to ML model instances. Other differences aside, these works do not address Kubernetes integration nor deal with monitoring cost reduction.

%% file: sources/system.tex
\section{System Design}
\label{sec:system}
\subsection{Architectural Elements and Deployment Model}
\begin{figure*}[t]
	\centering
	\includegraphics[width=\textwidth]{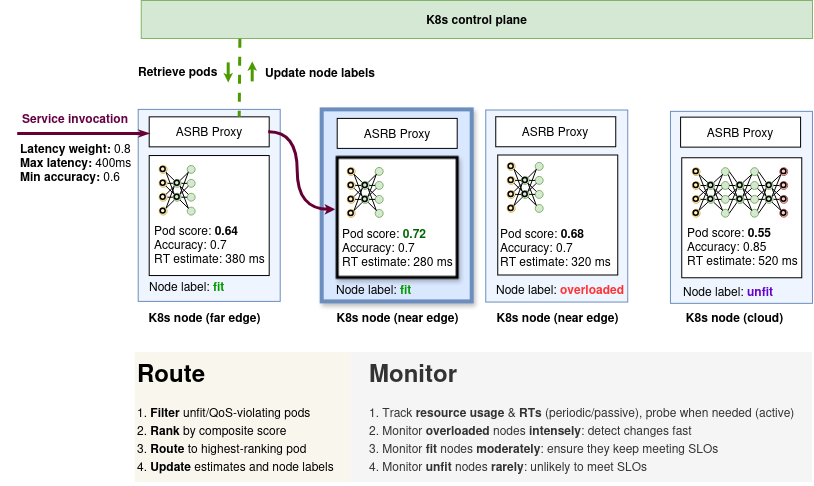}
	\caption{System architecture and operatations.}
	\label{fig:arch}
\end{figure*}

The deployment model we assume involves \emph{clients}, such as IoT devices, user applications, or other co-existing services, directing service invocation requests (typically, HTTP calls to application API endpoints, such as to an image classification service) to \emph{routing proxies}. Each proxy independently decides on a per-request basis on the most appropriate \emph{service instance} to forward the request to, out of a number of candidate instances deployed over a cluster of compute nodes managed by Kubernetes, the de facto framework for orchestrating containerized applications. Nodes may span cloud VMs and edge devices. 
Our approach follows the QEdgeProxy design introduced in our prior work~\cite{vcilic2024qedgeproxy}. We adopt the way it integrates with K8s and build on its code base to implement ASRB. However, we depart from its latency-centric design by supporting multi-dimensional, ML-oriented QoS objectives, and reduce its monitoring cost. \figurename~\ref{fig:arch} provides an overview of our system design and functionality.

\noindent\textbf{Service Pods.}  
We focus on ML inference serving, where ML models for a specific task (e.g., image classification) 
are deployed as containers over a K8s cluster. These service instances (pods, in K8s terminology) may run different models that are functionally equivalent, i.e., capable of handling the same task and accepting the same input over a unified interface, but exhibiting different performance characteristics and resource requirements. Information about these characteristics (e.g., expected ML prediction accuracy) is encoded in instance metadata. 
Note that instance placement decisions are beyond the responsibilities of our scheme.

\noindent\textbf{ASRB Node-Local Routing Proxy.}  
The ASRB proxy is deployed on each node as a \texttt{DaemonSet}, a Kubernetes construct ensuring that a copy of a pod runs on all (or select) nodes in the cluster. This deployment choice enables node-local request handling and avoids a centralized routing bottleneck. Different deployment options are possible: the service provider can bundle a distinct proxy with each service so that multiple proxies may co-exist on a single node, have one proxy per node responsible for multiple applications, or deploy proxies at specific edge nodes. The ASRB proxy is responsible for the following tasks: (i) \emph{scoring} of service pods using a scalar metric that jointly considers the latency and application-specific quality (in our case, ML model accuracy) each pod offers, (ii) dynamic pod discovery and health checking, (iii) instrumenting the adaptive monitoring of nodes, and (iv) fallback request routing when the initially preferred target is unavailable or does not satisfy current QoS constraints. It interfaces with the K8s control plane via the K8s API to retrieve lists of eligible pods and node/pod metadata, and updates internal caches via \emph{informers}, i.e., a K8s mechanism that allows to watch for resource updates. Pod discovery, score computation, and monitoring are executed asynchronously. Notably, ASRB dynamically assigns quality attributes to nodes via K8s \emph{labels}: each node receives a continuously updated label that characterizes its suitability for serving requests.

\noindent\textbf{Monitoring Subsystem.}  
ASRB tracks both resource-oriented metrics (CPU/memory use from K8s Metrics API) and latency-related data (by recording the latency of HTTP probes and observing response times), introducing a low-overhead \emph{adaptive monitoring strategy} (\S\,\ref{sec:monitoring}).
Monitoring information is made available through a metrics endpoint, which is consumed by Prometheus~\cite{Rabenstein15prometheus}, a widely used monitoring system in K8s-based environments.

\subsection{Node Labeling Mechanism}
We use Kubernetes labels, i.e., lightweight key--value metadata that we attach to each node and update based on observations made by proxies. A node may be labeled \emph{fit}, \emph{unfit}, or \emph{overloaded}, where \emph{fit} and \emph{unfit} reflect the aggregated score derived from latency and model accuracy, and \emph{overloaded} indicates excessive CPU or memory utilization. Labels are assigned by proxies independently via calls to the K8s API and are maintained at the Kubernetes control plane level, allowing proxies to indirectly share node state (e.g., an observation that a node is overloaded) and guide subsequent routing and monitoring decisions, as we will describe in \S\,\ref{sec:score-based-routing} and \S\,\ref{sec:monitoring}, respectively. We note that while maintaining labels has reasonable memory cost even for large numbers of nodes, the effects of centrally updating this state via the K8s API at scale require further study.

\section{Request Routing Logic}
\label{sec:score-based-routing}
\subsection{Scoring function design}
ASRB routes requests based on a weighted scoring mechanism that combines latency and model accuracy into a unified QoS metric. For each pod, the proxy collects the most recent latency measurement and retrieves the model's accuracy annotation.\footnote{ML model accuracy information is assumed to be known \emph{a priori} and available externally (e.g., by the application provider) as a metadata attribute, i.e., it is not measured at runtime by ASRB, contrary to latency. However, if the use case allows, the service provider may implement mechanisms to monitor accuracy and update these metadata at runtime.}
The balance between these two factors is determined by the \emph{latency importance weight} parameter $\lambda \in [0,1]$, which is supplied with each request.
The scoring function is therefore defined as:
\begin{equation}
	S_p = \lambda \cdot \left(1 - \min\left(\frac{L_p}{1000}, 1\right)\right)
	+ (1 - \lambda) \cdot A_p
	\label{eq:score}
\end{equation}

\noindent where $\lambda$ controls how strongly routing favors low latency, $L_p$ is the most recent latency measurement for pod $p$, and $A_p$ is a static model-specific accuracy value. Latency is transformed into a normalized latency score in $[0,1]$. The cap at 1000\,ms bounds the influence of outliers.
Notably, it is up to service providers to configure $\lambda$ and thus drive request routing towards the appropriate trade-off point, capturing service provider QoS priorities that may vary for different applications. For instance, in interactive user-facing applications, a higher $\lambda$ prioritizes responsiveness, while in classification-critical systems, a lower $\lambda$ increases the weight of model performance in the selection process. 
ASRB does not prescribe \emph{how} $\lambda$ is selected. For example, service providers can observe Service Level Objective (SLO) fulfillment performance, both historically and online, and use it as feedback to \emph{learn} to dynamically adjust $\lambda$.
We also remark that other quality objectives are straightforward to support. For example, tracking failed requests per instance can be used to maintain a respective instance availability metric, and weighted decisions that also factor in service availability could be made in an identical way. 

\subsection{Score-Based Routing Algorithm}
A proxy executes a lightweight routing algorithm for each incoming request, which selects the most suitable service instance. The routing process is summarized in Algorithm~\ref{alg:asrb-routing}. 

Specifically, ASRB retrieves all pods for the requested service from the K8s API and retains only those in the \texttt{Ready} state. For each ready pod, it determines the label of its hosting node (\texttt{fit}, \texttt{overloaded}, or \texttt{unfit}), and obtains its most recent latency estimate (from the local cache or via latency approximation if the value is stale), as well as its model accuracy annotation. It proceeds by computing the QoS score for each pod using Eq.~(\ref{eq:score}), extracting the value of $\lambda$ from the request. 

At this point, ASRB selects pods running on nodes labeled \texttt{fit} and discards those whose score falls below a configured minimum threshold. If no suitable \texttt{fit}-node pods are available,\footnote{It is also possible that a high-ranking pod fails to meet a constraint. This can happen if, e.g., ASRB is configured to prioritize for accuracy, but at the same time there is a stringent response time constraint that should be met.} it evaluates pods on \texttt{overloaded} nodes. If neither \texttt{fit} nor \texttt{overloaded} nodes provide a pod with an acceptable score, ASRB may optionally fall back to pods on \texttt{unfit} nodes or revert to a simple strategy such as random selection among healthy pods. Finally, it forwards the request to the highest-scoring pod from the final candidate set. Algorithm~\ref{alg:asrb-routing} performs a fixed number of constant-time operations per pod hosting the requested service (cache lookup and score computation take $O(1)$ time and ranking evaluates each pod at most once), and therefore routes a request in time linear in the number of pods discovered.
	
After a response is received, ASRB updates the latency measurement for the selected pod's host, recomputes the best pod score per node, and uses this score and resource usage to update the selected node's label for future routing and monitoring decisions. 
\begin{algorithm}
	\caption{ASRB Score-Based Request Routing Algorithm}
	\label{alg:asrb-routing}
	\begin{algorithmic}
	\normalfont
	\Require{Incoming request $r$, latency weight $\lambda$}
	\Ensure{Selected pod $p^*$}
	
	\State \textbf{(1) Service Discovery:} Retrieve all \texttt{Ready} pods for service $r$.
	
	\State \textbf{(2) Pod Categorization:}
	\begin{enumerate*}[label=\alph*)]
		\item collect recent latency $L_p$ from cache,
		\item collect score $S_p$ from cache,
		\item classify pods into:
		\textit{bestPods}: healthy and not overloaded,
		and \textit{overloadedPods}: healthy but above resource limits.
	\end{enumerate*}
	
	\State \textbf{(3) Latency Refresh (optional):}
	If too few pods have valid latency data, the proxy triggers an asynchronous latency-refresh routine to update stale or missing latency information, as summarized in Algorithm~\ref{alg:latency-refresh}.
	
	\State \textbf{(4) Score Computation:}
	For each candidate pod $p$:
    \vspace{-3mm}
	\[
	S_p = \lambda \cdot (1 - \min(L_p/1000, 1)) + (1 - \lambda) \cdot A_p
	\]
    
    \vspace{-3mm}
	\noindent where $A_p$ is the model accuracy from annotations.
	
	\State \textbf{(5) Ranking:} Select the highest-scoring pod within:
	\begin{enumerate*}[label=\alph*)]
		\item \textit{bestPods}
		\item else \textit{overloadedPods}
		\item else \textit{random from all healthy pods}
	\end{enumerate*}
	
	\State \textbf{(6) Forwarding:} Forward request $r$ to the selected pod $p^*$.
	
	\State \textbf{(7) Feedback Update:}
	\begin{enumerate*}[label=\alph*)]
		\item measure RTT,
		\item update latency estimate and pod score,
		\item update node QoS label (\texttt{fit}/\texttt{unfit}).
	\end{enumerate*}
	
	\State Note: The overloaded label is not set in this feedback step; it is maintained separately by the resource-monitoring component based on CPU and memory thresholds.
\end{algorithmic}	
\end{algorithm}

\begin{algorithm}
	\caption{Asynchronous latency refresh routine}
	\label{alg:latency-refresh}
	\begin{algorithmic}
	\normalfont
	\Require{Set of candidate pods $\mathcal{P}$}
	\Ensure{Updated latency cache entries and recomputed scores}
	
	\ForAll {$p \in \mathcal{P}$ }
		\If{latency for $p$ is missing or outdated}
			\State send lightweight request to $p$ at \texttt{/echo}\;
			\State measure round-trip time $L_p$\;
			\State update latency cache for $p$ with $L_p$\;
			\State recompute score $S_p$ using $L_p$ \& cached accuracy $A_p$\;
			\State store updated $S_p$ in the score cache\;
		\EndIf
	\EndFor
	\end{algorithmic}
\end{algorithm}

\vspace{-4mm}
\section{Adaptive Monitoring}
\label{sec:monitoring}
\subsection{Design Principles}
ASRB needs to keep an up-to-date view of the infrastructure and service state, to make quality-informed routing decisions. Aggressive monitoring helps maintain accurate state, which is important for high-quality routing decisions, but the overhead can be significant, particularly for large-scale deployments widely distributed over the continuum. To address this issue, our monitoring mechanism design is driven by the following intuition: On the one hand, static, fixed-interval monitoring may waste resources by probing nodes unnecessarily when conditions are stable, or miss important changes when the update frequency is too low under volatile conditions. On the other hand, there are nodes that, due to physical distance or the fact that they host low-quality ML models, may fail to meet the latency or accuracy QoS thresholds put in place by the service provider; such nodes are likely to be filtered out in routing decisions and become less relevant for intense monitoring. The system can therefore \emph{allocate more observation effort to uncertain or critical nodes, and reduce monitoring for consistently stable or obviously unsuitable ones.} This boils down to the following monitoring principles:
\begin{itemize}
	\item \texttt{Fit} or \texttt{overloaded} nodes receive higher monitoring frequency, as they are more likely to be selected for routing.
	\item Monitoring frequency is increased for nodes with volatile scores or borderline performance (e.g., latency spikes, occasional failures).
	\item Nodes with consistent performance over a defined window are monitored less frequently to conserve resources.
	\item Nodes labeled \texttt{unfit} are still refreshed periodically, but at the lowest frequency, allowing the system to detect recovery without generating unnecessary overhead.
\end{itemize}

\subsection{Adaptive Active-Passive Monitoring Strategy}
ASRB monitors latency through two mechanisms: \emph{passive}, where feedback updates are collected after each request, and \emph{active}, which initiates on-demand approximation when cached values become stale. 
In this case, the proxy issues a lightweight request to the pod's \texttt{/echo} endpoint to estimate the current round-trip time. These active probes are triggered adaptively and only when required, rather than on every request.
When a real latency measurement becomes available after a period of approximation, the cached latency value is updated to maintain an exponentially weighted moving average. This prevents abrupt score changes. Latency approximation is regulated by a cooldown mechanism. After an approximation cycle is triggered for a service, further approximation attempts are suppressed for a configurable interval to avoid excessive probing. During this interval, cached latency estimates are reused.

Resource usage metrics, including indicators such as CPU utilization, memory usage, and pod health status, are exposed by K8s and are obtained through the Metrics API or via Prometheus. They are refreshed periodically and used to label nodes as \texttt{overloaded} when predefined thresholds are exceeded. These labels are  used in two ways: (i) they steer the monitoring process by refreshing \texttt{overloaded} nodes more frequently than \texttt{fit} nodes, which are in turn refreshed more frequently than \texttt{unfit} ones, and (ii) they support routing by treating pods on \texttt{overloaded} nodes as a fallback option when no suitable pods on \texttt{fit} nodes are available.

%% file: sources/evaluation.tex
\section{Evaluation}
\label{sec:eval}
\subsection{Experimental Setup}
We evaluate our scheme against mechanisms from the state of the art on a k3s-based cluster, serving an image classification task with a number of pods hosting either the MobileNet-V2 (faster, lower accuracy) or the RestNet-50 model (slower, higher accuracy). We generate a stream of 1200 service requests to the exposed API endpoint, and the experiment consists of six phases (T1--T6) where the setup undergoes controlled topology changes, overload events, and recovery periods. Our cluster forms a representative (simulated) far edge (IoT device space) -- near edge (telco/MEC data center) -- cloud topology. The following compute nodes, each running as a separate Ubuntu 22.04 virtual machine in a local data center, are deployed over these three tiers: \textbf{(Master node)} A cloud-tier controller running the K8s control plane and hosting inference pods executing both MobileNet and ResNet models, reachable from the IoT device (request source) with a 150\,ms RTT; \textbf{(Worker~1 and Worker~2)} stable near-edge K8s nodes reachable at a 50\,ms RTT from the request source, hosting both MobileNet and ResNet inference pods; \textbf{(Worker~3)} on-premise far-edge K8s node, reachable at a 5\,ms RTT; initially without a pod (T1), receives a MobileNet pod in T4, and becomes overloaded in T5; \textbf{(Worker~4)} far-edge K8s node reachable at a 10\,ms RTT, hosting only a MobileNet model; added in T2, rebooted in T3, and overloaded in T5. Latencies are controlled using the Linux \texttt{tc} utility, and each node runs a request router instance capable of executing our candidate routing strategies. The six experiment phases are as follows:
\begin{itemize}
	\item \textbf{T1 -- Initial topology:} Workers 1--3 active; Worker~3 has no model pod.
	\item \textbf{T2 -- Node arrival:} Worker~4 joins and becomes available.
	\item \textbf{T3 -- Reboot and recovery:} Worker~4 becomes unavailable, then recovers.
	\item \textbf{T4 -- Pod deployment:} Worker~3 receives a MobileNet pod.
	\item \textbf{T5 -- Overload:} Workers 3--4 are overloaded:
    a temporary pod is deployed on each, executing CPU and memory load using the \texttt{stress} utility (\texttt{stress --cpu 4 --vm 2 --vm-bytes 1G}).  
    This pushes node CPU utilization above 85\%, triggering 
	the overload condition.
	\item \textbf{T6 -- Recovery:} Overload removed.
\end{itemize}

The following routing strategies are evaluated: (i) \textbf{ASRB} with $\lambda \in \{1, 0.5, 0\}$, and under two different monitoring configurations; (ii) \textbf{QEdgeProxy (QEP)}~\cite{vcilic2024qedgeproxy} with its latency-only routing and static monitoring configuration; (iii) two different configurations of \textbf{proxy-mity}~\cite{fahs2019proxy}, corresponding to pure proximity-based routing ($\alpha=1$) and a version that trades latency for more even load balancing ($\alpha=0.8$); (iv) \textbf{Round Robin.}

\subsection{Response Time vs. ML Accuracy Performance}
\label{sec:eval-tradeoff}
\begin{figure}[t]
	\centering
	\includegraphics[width=\columnwidth]{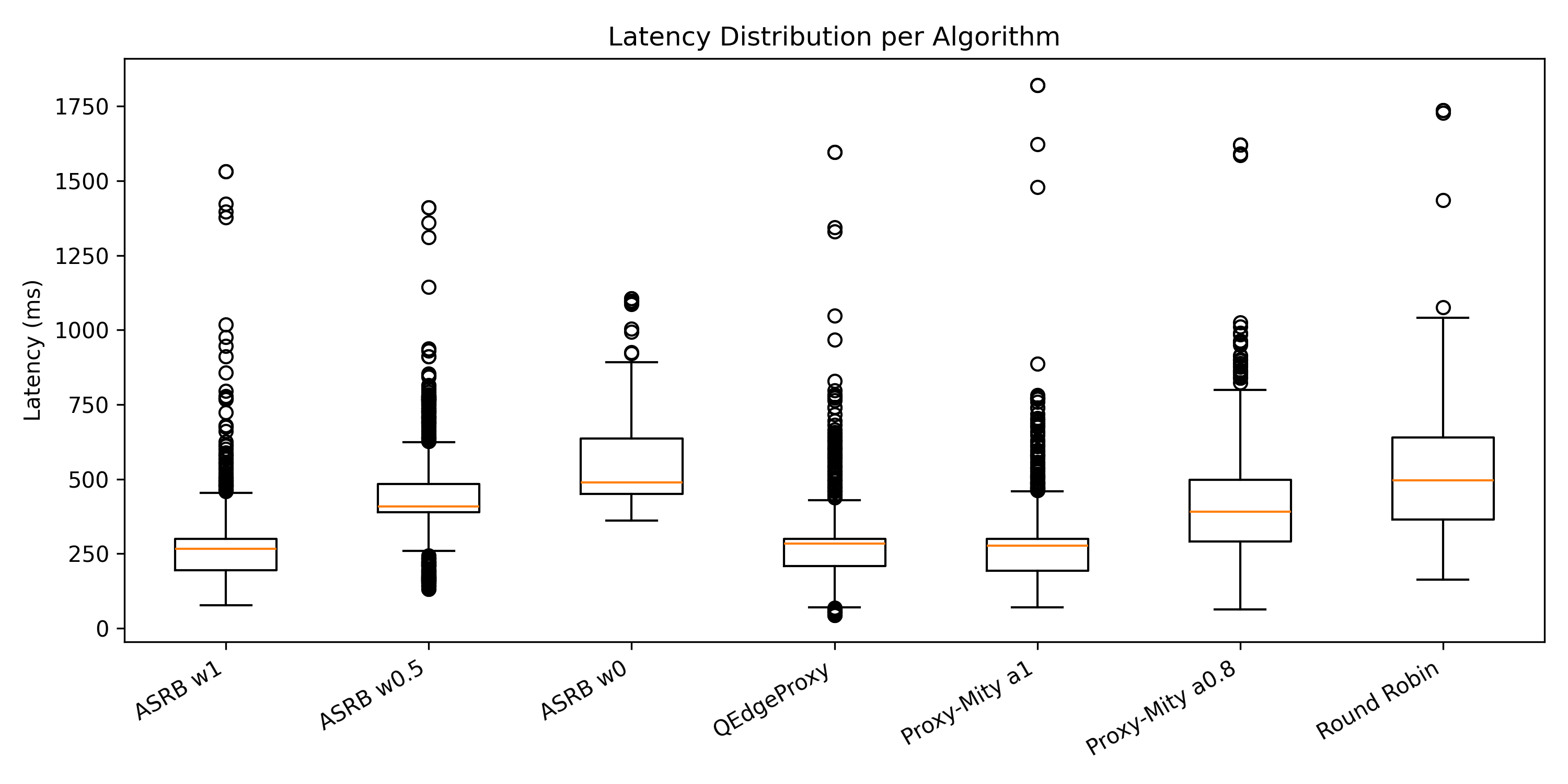}
	\caption{Latency distribution across all routing algorithms.}
	\label{fig:lat-boxplot}
\end{figure}
\figurename~\ref{fig:lat-boxplot} presents response time statistics for the different request routing schemes over all phases of the experiment.
ASRB with $\lambda = 1$ achieves not only the lowest median latency but also a small interquartile range, indicating a highly stable latency profile.

Although ASRB~($\lambda=1$), QEP, and proxy-mity are all latency-oriented, they differ in how latency-based decisions are maintained under dynamic conditions.
Proxy-mity relies on network latency ranking without explicit node-state tracking. Consequently, it reacts only to instantaneous latency values and does not handle transient degradation. QEP incorporates latency feedback and resource checks, but applies them in an instantaneous manner. Overload decisions are based on momentary CPU or memory thresholds and are not persisted through explicit node-state labels or cooldown periods. In addition, pods may be treated as routable before networking information (e.g., \texttt{PodIP}, \texttt{HostIP}) is fully initialized, which leads to failed requests during node joins or reboots. ASRB~($\lambda = 1$) preserves a latency-centric objective but stabilizes routing through explicit node labeling, cooldown logic, and stricter pod admission semantics. Under dynamic conditions, these mechanisms reduce latency variance and extreme outliers by preventing routing to temporarily degraded or not-yet-ready nodes. While the overall latency differences between the latency-oriented approaches remain small, they are statistically significant (see Table~\ref{tab:stat-sig}).

\begin{table*}[htbp]
\centering
\caption{Mean pairwise response time difference ($\overline{\Delta}$) vs. baselines in latency-oriented configurations. Negative values indicate lower response time than the baseline. The null hypothesis $H_0: \overline{\Delta} = 0$ (no latency difference) is rejected at the 95\% level: Phase-stratified bootstrap CIs (computed using the BCa method~\cite{efron2016computer} and including 100\,000 resamples) exclude 0; we also report the results of one-sided t-tests for the directional alternative hypothesis $H_1: \overline{\Delta} < 0$.}
\label{tab:stat-sig}
\begin{tabular}{lllll}
\hline
\textbf{Baseline} & $\mathbf{\overline{\Delta}}$ \textbf{(ms)} & \textbf{95\% Bootstrap (BCa) CI} & $\mathbf{t}$ & $\mathbf{p}$ \\
\hline
proxy-mity & -9.79 & [-18.76,-0.83] &      -2.07 & .0193   \\
QEP & -18.64 & [-26.62,-9.98] &      -4.22 & $<$.0001  \\
\hline
\end{tabular}
\end{table*}

At the same time, ASRB allows to flexibly address latency-accuracy trade-offs: For example, when configured with $\lambda = 0.5$, while showing a noticeably higher median latency (440\,ms) and increased variability compared to a pure latency-oriented configuration, \emph{it routes 85.4\% of the requests to high-accuracy instances} (vs. 7.5\% for $\lambda=1$, 7.1\% for proxy-mity, and 2.6\% for QEP), selecting ResNet50 preferentially, while still routing to MobileNet during latency spikes or transient overloads.
At the other end of the configuration space, ASRB~($\lambda = 0$) exhibits a markedly higher median latency and an expanded upper tail, reflecting the slower inference time of the ResNet50 model, which it selects almost exclusively for higher accuracy.

\subsection{Responsiveness to Environment Dynamics}
\label{sec:eval-responsiveness}
\begin{figure}[t]
	\centering
	\includegraphics[width=\columnwidth]{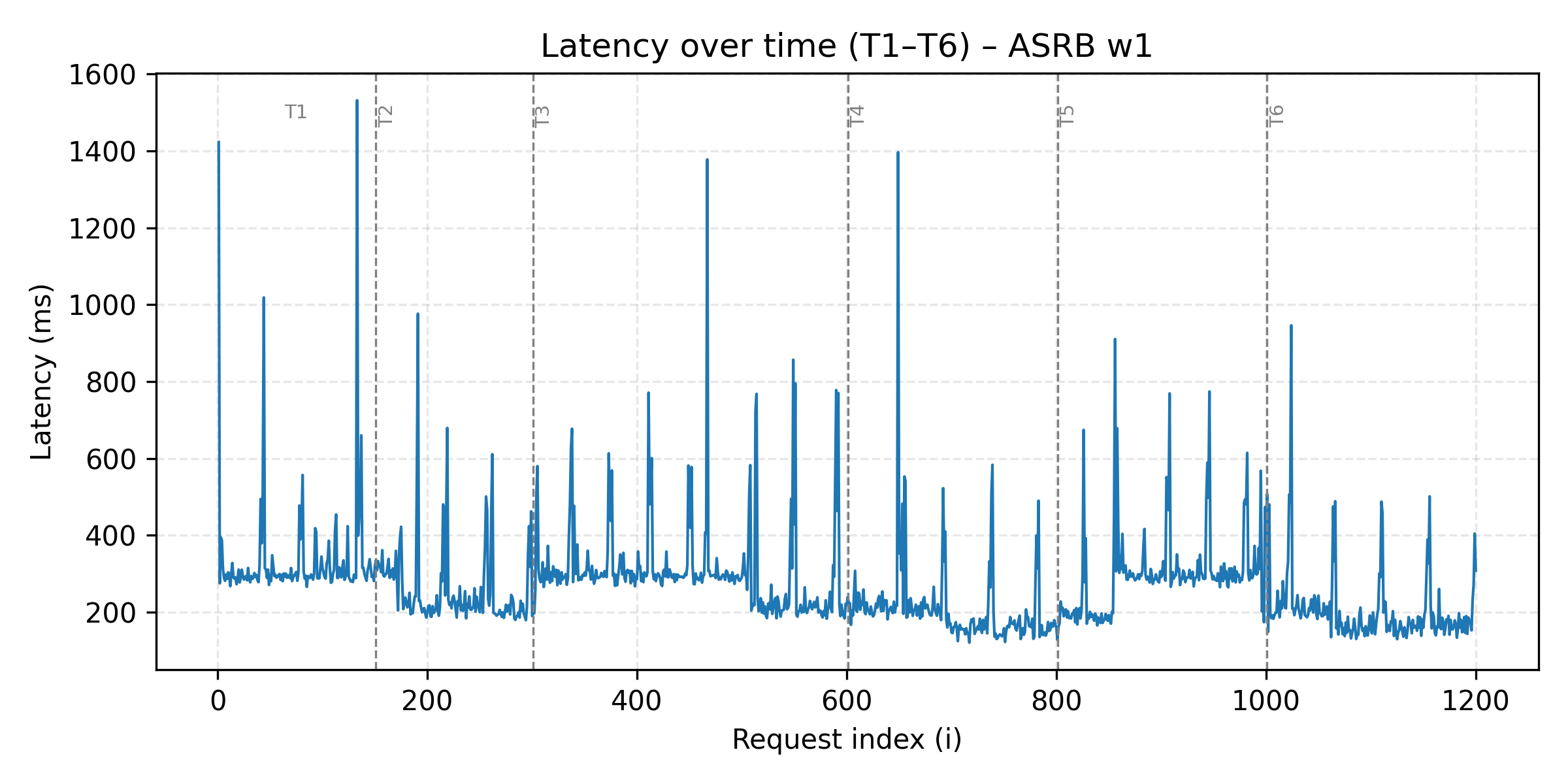}
	\caption{Latency evolution across T1--T6 for ASRB ($\lambda=1$).}
	\label{fig:dyn-asrb}
\end{figure}
\figurename~\ref{fig:dyn-asrb} shows the request-by-request latency evolution over the entire 1200-request experiment for ASRB ($\lambda = 1$). Compared with the baselines (figures omitted in the interest of space), this ASRB configuration achieves the most stable profile across all phases. During T1, it consistently chooses the fastest nodes (Worker~3 or Worker~1), resulting in low latency with only minor spikes caused by transient network fluctuations. When Worker~4 joins in T2, ASRB incorporates it almost immediately, reflected by a small dip in average latency. The reboot event in T3 is also handled quickly: the latency spikes are short, and the system reroutes within a few requests. In T5, ASRB detects the overload on Workers~3--4 early and shifts traffic toward stable nodes, preventing the long clusters of high latency experienced by the baselines. T6 returns to a clean and stable profile.

QEP shows similar trends in stable phases but is slower to detect the rebooted node in T3, while it continues sending requests to degraded nodes for noticeably longer period than ASRB in T5. This confirms that ASRB's adaptive monitoring is more suitable for dynamic edge environments. Similar behavior was observed for proxy-mity, while Round-Robin shows high variance even in stable phases due to its static nature. 
It should finally be noted that during the whole duration of the experiment, ASRB experienced only 2 transient failures (vs. 36 for QEP).

\subsection{Monitoring Cost Reduction}
\label{sec:eval-monitoring}
We compare the monitoring cost of ASRB ($\lambda=1$) and QEP using the following metrics: (i) number of monitoring API calls to the K8s control plane, and (ii) volume of the associated monitoring traffic. QEP polls for information about \emph{all} nodes and instances at fixed intervals, whereas ASRB  controls polling frequency adaptively, resulting in fewer such calls. On the flip side, ASRB needs to send PATCH requests to the K8s control plane to update pod scores and node labels, which add to its monitoring cost. We evaluate different monitoring configurations, parameterized by the following: (i) node state monitoring interval (node metrics cache time), and (ii) node label update interval. For ASRB, (i) represents the monitoring period for overloaded nodes; fit/unfit nodes are monitored at $2\times$\,/\,$3\times$ that interval. For QEP, all nodes are monitored at this interval. Note that (ii) is only relevant for ASRB, and a PATCH API call to update a node label takes place only if the label would change. Despite a potential increase in API calls, ASRB significantly reduces the \emph{volume} of monitoring traffic, as Table~\ref{tab:payloads} shows. This is because the monitoring overhead of the baseline is dominated by the much larger payload size per call: QEP repeatedly queries all workers regardless of their relevance or state. 

While the monitoring overhead in absolute terms is limited due to the small size of the testbed, the achieved savings (29\%-73\% less monitoring traffic than QEP for the same metric update interval) will be significant in large-scale deployments. At the same time, such savings come with only a modest latency penalty in \emph{some} configurations. For example, 
with metric and label update intervals of 15\,s and 60\,s, respectively, \emph{ASRB saved 73\% of monitoring traffic at the expense of only 5.3\% higher mean response time than QEP}, while in its most aggressive monitoring configuration, \emph{ASRB's mean response time was 6.7\% lower than that of QEP, still with 69\% less monitoring traffic}.

\begin{table*}[t]
\setlength{\tabcolsep}{4.5pt}
	\centering
	\caption{Monitoring overhead for different configurations. We also report percentage traffic savings and the mean pairwise response time difference ($\overline{\Delta}$) vs. the respective QEP configuration. For configuration 15\,s/60\,s, the null hypothesis of ASRB being inferior (slower) than QEP by a margin higher than 10\% is rejected at the 95\% level: The phase-stratified bootstrap CI ($[-23.29, -2.80]$) excludes 0 and a one-sided t-test yields $t=-2.38, p = .0086$. Similarly, the non-inferiority margin for configuration 60\,s/60\,s is 3\%. 
    Across the board, ASRB reduces monitoring traffic by 29\%-73\%.} 
	\label{tab:payloads}
	\begin{tabular}{lccccc}
		\hline 
		& \multicolumn{2}{c}{\textbf{Update intervals}} 
		& \textbf{API calls} 
		& \textbf{Traffic (MB)}
        & \textbf{$\mathbf{\overline{\Delta}}$ \textbf{(ms)}}\\
		& \makecell{\textbf{Metrics}} 
		& \makecell{\textbf{Labels}} 
		& & & \\
		\hline
		
		\multirow{2}{*}{QEP}
		& 15\,s & --  & 136 & 3.29 & \\
		& 60\,s & --  & 40  & 1.07 & \\
		
		\hline
		
		\multirow{4}{*}{ASRB}
		& 15\,s & 5\,s  & 85 & 1.01 (-69\%) & -18.64 (-6.7\%)\\
		& 15\,s & 60\,s & 73 & 0.89 (-73\%) & +14.94 (+5.3\%)\\
		& 60\,s & 5\,s  & 81 & 0.76 (-29\%) & -16.62 (-5.6\%)\\
		& 60\,s & 60\,s & 72 & 0.63 (-41\%) & -1.26 (-0.4\%)\\
		
		\hline
	\end{tabular}
\end{table*}

%% file: sources/conclusion.tex
\section{Conclusion}
\label{sec:conclusion}
We presented ASRB, a decentralized request routing scheme tailored to inference serving, capable of balancing workloads while considering multi-dimensional QoS objectives---particularly latency and prediction quality---in a flexible way. ASRB integrates natively with Kubernetes ecosystems and enables IoT service providers to prioritize conflicting quality criteria depending on the specific requirements of applications that rely on ML service instances across the computing continuum. Its decentralized design and reduced monitoring overhead make it suitable for dispatching workloads in large-scale distributed deployments. More sophisticated monitoring adaptations, studying the effects of node state updates in very large deployments,  and applying ideas from reinforcement learning to aspects such as dynamically adjusting QoS priorities, are directions for future work.